# AutoEIS: automated Bayesian model selection and analysis for electrochemical impedance spectroscopy

*Runze Zhang[a], Robert Black[b], Debashish Sur[c,d], Parisa Karimi[b], Kangming Li[a], Brian DeCost[e], John R. Scully[c,d], Jason Hattrick-Simpers[a*]*

a. Department of Material Science and Engineering, University of Toronto, Toronto, Ontario, Canada

b. Research Center of Energy, Mining, and Environment, National Research Council Canada, Mississauga, Ontario, Canada

c. Center for Electrochemical Science and Engineering, University of Virginia, Charlottesville, VA USA 22904

d. Department of Materials Science and Engineering, University of Virginia, Charlottesville, VA USA 22904

e. Material Measurement Laboratory, National Institute of Standards and Technology, Gaithersburg, MD, USA

Email: jason.hattrick.simpers@utoronto.ca

**Keywords:** Accelerated Material Discovery, Automated Analysis Technique, Electrochemical Impedance Spectroscopy, Bayesian inference, Evolutionary Algorithms

**Abstract:** Electrochemical Impedance Spectroscopy (EIS) is a powerful tool for electrochemical

analysis; however, its data can be challenging to interpret. Here, we introduce a new open-source tool named AutoEIS that assists EIS analysis by automatically proposing statistically plausible equivalent circuit models (ECMs). AutoEIS does this without requiring an exhaustive mechanistic understanding of the electrochemical systems. We demonstrate the generalizability of AutoEIS by using it to analyze EIS datasets from three distinct electrochemical systems, including thin-film oxygen evolution reaction (OER) electrocatalysis, corrosion of self-healing multi-principal components alloys, and a carbon dioxide reduction electrolyzer device. In each case, AutoEIS identified competitive or in some cases superior ECMs to those recommended by experts and provided statistical indicators of the preferred solution. The results demonstrated AutoEIS's capability to facilitate EIS analysis without expert labels while diminishing user bias in a high-throughput manner. AutoEIS provides a generalized automated approach to facilitate EIS analysis spanning a broad suite of electrochemical applications with minimal prior knowledge of the system required. This tool holds great potential in improving the efficiency, accuracy, and ease of EIS analysis and thus creates an avenue to the widespread use of EIS in accelerating the development of new electrochemical materials and devices.

1. **Introduction:**

There is a pressing environmental and societal need for novel electrochemical materials for deployment as anodes/cathodes for batteries [1], alloys that resist corrosion and oxidation in harsh environments [2], electrocatalysts for fuel cells [3], oxygen evolution reaction (OER) [4], hydrogen evolution reaction [5], and $CO_2$ reduction reactions [6]. Over the past 10 years programs such as the Materials Genome Initiative have sought to meet this need by combining computation, experimentation, and data science to reduce the time required to deploy new materials to 5 years - 10 years down from the current 10 years - 20 years [7–12]. The release of the Mission Innovation report on Materials Acceleration Platforms posited that by combining experimental and computational automation with data science and artificial intelligence (AI), self-driving labs could be created to increase the rate of materials discovery by 2 to 3 orders of magnitude [13]. Towards this aim, the development of automated analysis techniques using machine learning serves as a crucial piece. Specifically, such tools can process vast amounts of data generated by self-driving labs in a high-throughput manner, enabling less human intervention and reduced bias in decision-making. AI augment data analysis is necessary for the realization of this new paradigm in material discovery.

In the pursuit of developing novel electrochemical materials, electrochemical impedance spectroscopy (EIS) is a cornerstone analysis technique for a wide range of electrochemical systems [14]. It is regularly used to investigate the mechanisms and to extract key parameters of electrochemical systems, such as to study the kinetics of corrosion and extract the coating's dielectric properties [15–18]. At its core, the basic principle of EIS involves applying a small alternating current (AC) potential or current perturbation to an electrochemical system and measuring the corresponding current or voltage response to get the impedance of the system at different frequencies. The resulting impedance values are complex quantities that encompass both the reactance and resistances of the electrochemical phenomena as a function of AC frequency and are typically presented as the complex plane plot (called Nyquist plot or Cole-Cole plot) or phase-magnitude plot (called Bode plot) [17]. Through common techniques such as equivalent circuit model (ECM) fitting, the impedance spectra provide a systematic method of separating out the different electrochemical processes occurring in the system based on their responses at different frequencies [19–21]. However, performing quantitative EIS analysis is non-trivial, as it relies largely on expert knowledge, and can be highly susceptible to the injection of human bias into its analysis [22].

The ECM method is by far the most common and widely accepted approach to analyzing and drawing physical interpretation from impedance spectra [20,23]. In the ECM method, all the electrochemical reactions are represented by an electric circuit composed of a set of basic circuit components, such as resistors, capacitors, constant phase elements (CPEs), and inductors arranged in either a parallel or series. By modeling impedance spectra in this way, analysts can interpret physical processes occurring in the electrochemical system through analogy to circuit components and their values. Unfortunately, generating ECMs is quite subjective and can be prone to user bias. Further, there is no guaranteed unique solution for a given impedance spectra [17,21,24]. To mitigate this degeneracy, the EIS community has developed a set of knowledge-based heuristics and proposed specialized ECM elements such as Warburg elements and Gerischer elements to constrain the potential solution space [25,26]. Nevertheless, this neither prevents routine overfitting of EIS data—even poor-quality data can still be adequately fitted by adding enough circuit elements—nor ensures a robust search of the solution space [27] The identification of the optimal ECM, *i.e.,* the simplest model that correctly reflects the system's information, remains challenging.

Alternatively, the application of machine learning offers an approach to exploring the ECM solution space that can be more transparent and reproducible [28–37]. Evolutionary algorithms have been used since 2004 to explore ECM structure and have shown great promise in ECMs-based EIS analysis [28,36,37]. By combining genetic algorithm (GA) [38] and gene expression programming (GEP) [39], the evolutionary algorithms can automatically optimize both the circuit structure and fitting parameters of circuit components by sequentially searching through the solution space.

Evolutionary algorithms provide an unconstrained path to objectively exploring the candidate ECMs, however, the non-uniqueness of the solutions yields merely a probabilistic sampling of candidate ECMs with sufficient quality. Any given iteration is as likely to produce a non-realistic solution as a physically plausible solution. Furthermore, the objective function used in evolutionary algorithms to search for the optimal ECMs only relates to the fitting quality, which is usually mean square error (MSE) or mean absolute percentage error (MAPE). These metrics are insufficient and ineffective for ECM selection since using fit quality alone tends to prioritize overfitted models. Worse, these metrics are not strictly followed by the community during model selection, which still mainly relies on researchers' intuition. Within the community, circuits with non-optimal fits that conform to community standards are sometimes preferred to optimal fits that violate community standards. There is then a misalignment between the purely algorithmic MSE-driven evolutionary algorithm-generated ECMs and the softer human heuristic approach to fitting EIS. Therefore, to realize truly trustworthy automatic EIS analysis, it is of great importance to create an effective ECM evaluation strategy that helps to identify and reject non-realistic models and suggests physically plausible models.

Bayesian inference[40], an iterative analysis tool, is a complementary tool to evolutionary algorithms because of its ability to provide statistical information about the parameter values. Recently, there has been increasing interest in utilizing Bayesian approaches to assist with EIS analysis [41–46]. To date, none of them involve inferring the values of the ECMs in light of the EIS data. In the case of ECMs, Bayesian inference provides probability distributions of electric components contained in each ECM according to a given EIS data. The probability distributions revealed by Bayesian inference provide a window into the plausibility of the overall model, which can be used to guide and justify ECM selection.

In this work, we combine evolutionary algorithms with Bayesian inference, to create a scientific AI that facilitates EIS analysis. The tool is available as an open-source Python-language-based package named AutoEIS. AutoEIS provides a generalized approach to automatically constructing reasonable ECMs without requiring an exhaustive mechanistic understanding of the underlying electrochemical systems. It objectively searches for potential ECMs and uses Bayesian inference to identify statistically plausible ECMs, greatly lowering the barrier to performing EIS analysis. To assess the algorithm's robustness and generalizability, we have tested AutoEIS on EIS datasets collected from three distinct electrochemical systems: oxygen evolution reaction (OER), steady-state passivation of alloys, and $CO_2$ reduction electrolyzer devices. In all cases, AutoEIS either successfully recommended new physically plausible ECMs or corroborated the expert's previously proposed solution. The success of AutoEIS across multiple electrochemical systems considered highlights its potential as a universal tool in accelerating electrochemical analysis for a range of topics including: electrocatalysts, investigations of corrosion mechanisms, etc.

## 2. Methodology

AutoEIS contains the following steps: data pre-processing, ECM generation, circuit post-filtering, and Bayesian inference. Its inputs are solely the impedance data and corresponding frequencies. Prior to analysis, the quality of the EIS data is evaluated according to the Kramer-Kronig relations (KK relations) [47], which removes erroneous measurements. Then AutoEIS uses evolutionary algorithms to search for a set (50 to 100) of high-quality candidate ECMs [32]. Subsequently, AutoEIS uses physical rules and soft human heuristics to perform model down-selection to reduce the number of non-realistic circuits. Bayesian inference is performed on each candidate ECM to explore the probability distributions of components. Through evaluation of inference quality, posterior distributions, and the corresponding predictive plots, ECMs with redundant components and imprecise parameter estimates are removed, and statistically plausible models are suggested. The selected models are then ranked by their widely applicable information criterion (WAIC) values to prioritize less complex ECMs with sufficient fitting quality [48]. Through this process, the statistically optimal model will be identified and suggested to the user.

## 2.1 Data pre-processing

To minimize the impact caused by invalid EIS measurements, we use Kramer-Kronig validation (KK validation) [49] to filter the points that do not satisfy the linearity, causality, and stability criteria. Any points exhibiting deviations larger than a default threshold of 5 % are removed. This filtering threshold was selected empirically based on expert insights to yield the cleanest filtered data without compromising our ability to resolve electrochemical reactions. In addition, a high-frequency filter is applied to remove all data points with positive imaginary impedance ($Im(Z) > 0$) at high-frequency ranges. These data points are artifacts of the electrochemical set-up associated with cabling inductance or contact connections that are not of scientific interest and would normally be eliminated through careful consideration of the scanned frequency range.

## 2.2 ECM generation

ECM generation is realized via the open-source program developed by Van Haeverbeke et al [32]. The program is designed to either fit the parameters of given ECMs or objectively search for potential ECMs for EIS measurements by evolutionary algorithms. Here we use its latter function to explore ECM space. There are two main hyperparameters that impact the exploration of circuit structures. The first hyperparameter named *head* restricts the complexity of the circuit search space by constraining the maximum number of circuit components included in potential ECMs. The second hyperparameter named *cutoff* determines the degree of the circuit simplification that sequentially drops components included in each ECM and compares the incidental loss for redundant components removal. For all the EIS data analyzed in this paper, the head was set to 12, and the cutoff was set to 0.80 unless otherwise specified. This parameter combination provides a reasonable search space with up to 11 circuit components that are sufficient to describe the electrochemical systems considered here and avoids the consideration of overly complex models. To ensure the evolutionary algorithm covers a great variety of ECMs, the ECM generation process was performed one hundred times for each EIS. The process usually takes 2 hours to 10 hours using a typical laptop (8-core intel i7 with 16 GB RAM) according to EIS complexity. This process can be greatly accelerated through multi-threading.

## 2.3 Post-filtering

The ECMs generated contain a mixture of realistic and non-realistic solutions. To ensure the models to be evaluated by Bayesian inference are physically meaningful, here we incorporated two physical filtering rules related to ohmic resistance and capacitors into AutoEIS to perform the ECMs down-selection.

The ohmic resistance (typically also referred to as solution resistance) reflects the summation of the system's bulk resistance (solution and electrode interfaces) to the current and can be directly extracted from the real impedance value at the highest frequency. AutoEIS estimates the ohmic resistance by extracting the real impedance value of the high-frequency data point with the minimum phase angle after pre-processing, adding a 15% buffer to the estimated resistance to accommodate measurement and calculation errors. Any ECM without the ohmic resistor or with an incorrect ohmic resistance is dropped. Additionally, here we delete all ECMs with any capacitor element because capacitors are only suitable for ideal polarization processes [50] These are both configurable options and can be adjusted, removed, or augmented when deploying AutoEIS. This filtering step effectively drops non-physical solutions and shrinks the number of candidate ECMs from about 100 to only a handful of more realistic ECMs.

## 2.4 Bayesian inference

Bayesian inference is used to evaluate each ECM's performance on the given EIS. During this process, the prior beliefs of the values of ECM components, specified as the prior probability distributions $p(\theta)$, are reallocated and updated according to the likelihood function $p(D|\theta)$ as new observations $p(D)$ are made. The $\theta$ represents component values and $D$ represents the data. The resulting distributions are called the posterior distributions $p(\theta|D)$ and provide updated credibility of the component parameters. The schematic workflow of applying Bayesian inference on each ECM is shown in **Figure 1**, with the following steps:

1) Initialize the prior probability distributions on circuit components
2) Sample components values according to the assigned distributions
3) Calculate the simulated EIS data according to the ECM's function
4) Compare the simulated data with the original EIS data to evaluate the likelihood function
5) Update posterior distributions according to the priors and ECM's likelihood function
6) Repeat steps 2 to step 6 for a given number of iterations

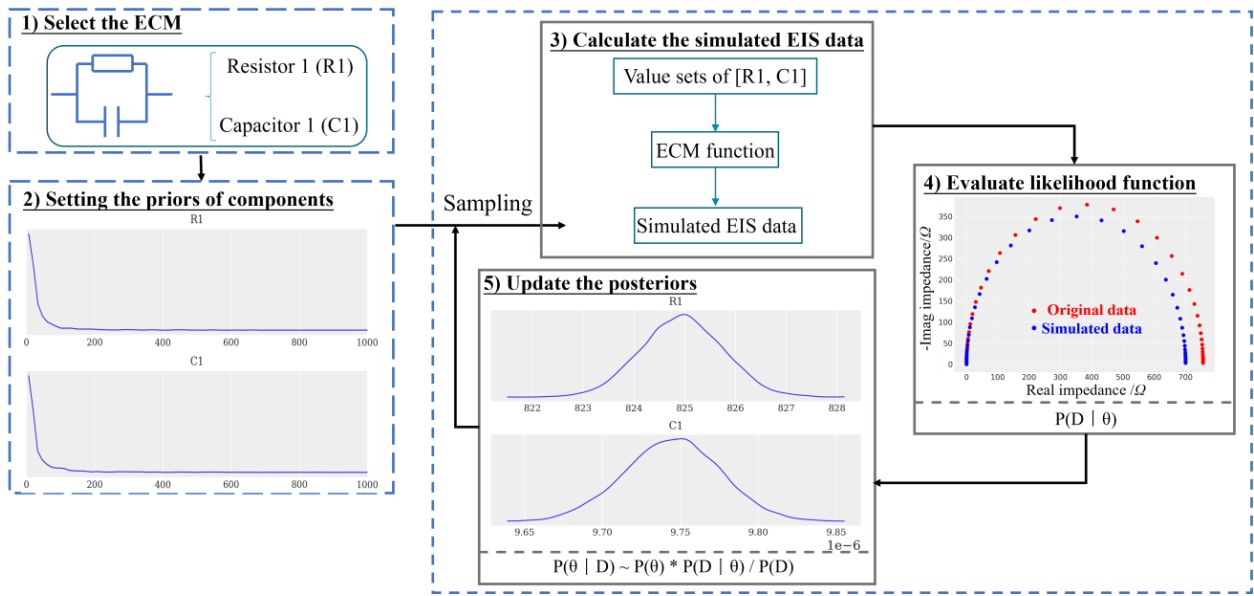

**Figure 1**. The schematic workflow of applying Bayesian inference on EIS analysis

The majority of ECMs generated by the evolutionary algorithm exhibit coefficients of determination ($R^2$) values higher than 0.99. Consequently, the evolutionary algorithm estimated values are employed to inform the prior settings. A weakly informative prior is selected via a scaling factor multiplied by the estimates – $LogNormal(2.5, 1.7) \times Estimated\ Values$, where 2.5 is the location parameter and 1.7 is the scale parameter of the distribution. It results in a prior where the samples are approximately centered around the estimated values and 99 % of the prior density lies within the range of (0.001 to 1000) times the estimated values. This diffuse prior prevents strong bias in the final estimates and ensures the data have a strong influence on the posterior distribution, while concentrating the prior distribution on roughly the right order of magnitude, which can vary substantially for ECM components. Also, the LogNormal distribution ensures all samples fall within the physically reasonable range of components since resistance, capacitance and inductance cannot be negative. All CPE alpha terms (represented as Pn)—the frequency-independent negative phase between current and voltage—that represent CPEs' degrees of similarity to perfect capacitors are assigned uniform distributions between 0 to 1, which encompasses their entire physically reasonable domain.

To mitigate the impact of sampling bias, a large sampling number of 10,000 with a warm-up of 1,000 steps was adopted, using the No-U-Turn sampling method [51] with a target acceptance probability of 0.8 for Bayesian inference. This sampling number permits sufficient iterations for obtaining stable distributions consistent with the EIS data. The Bayesian inferences typically require about 30 seconds per ECM (up to 3 minutes for the most complex ECM) using the laptop mentioned above (8-core intel i7 with 16 GB RAM). To enhance the model's robustness against noisy data, a HalfNormal probability distribution was employed for the error term between simulated and original EIS. This distribution was used to define the likelihood function, which assesses the goodness of fit between simulated and original EIS. The likelihood function can be represented as $\|Simulated\ EIS - Original\ EIS\| \sim HalfNormal(0, Error)$, where the residual between simulated and original EIS is calculated as a Euclidean distance and the standard deviation of the likelihood function is set as a HalfNormal probability distribution with 0 as its mean and *Error* as its scaling parameter. To quantify the convergence and repeatability of the Bayesian inference results on each ECM, the Bayesian inference process of each model is independently repeated four times for each ECM.

## 2.5  Model evaluation

**Figure 2** illustrates the algorithmic workflow of evaluating ECMs according to the Bayesian inference results. Though the evaluation strategy focuses on prioritizing the statistically plausible ECM, it retains suboptimal models with lower priority for subsequent expert inspection. Firstly, AutoEIS identifies ECMs with insufficient numerical explorations of the posterior distribution by examining numerical divergences ($N_d$) exhibited in posterior distributions. The occurrence of divergences indicates that the Markov chain has encountered regions of high curvature in the target distribution that may hinder adequate exploration of the posterior distribution [52–54]. The optimal value of $N_d$ is 0 and any inference with $N_d$ larger than 10 (0.1 % of the samples in the chain) is deemed a poor model here. The 0.1 % threshold selection was based on the consideration of the potential complex posterior geometry of ECMs. Through this step, ECMs with a $N_d$ value larger than 10 will be deprioritized.

Secondly, AutoEIS checks the posterior distributions of each ECM's components based on their shapes and highest density intervals (HDI), which describe and summarize the probability of estimated parameters falling within a specific value range [40] Ideally but not necessarily, each posterior distribution should exhibit a smooth distribution with a sharp, well-centered bell curve.

This is interpreted to indicate that the model is not overfitted, every component is irreplaceable, and that there is high confidence the element values fall within a narrow credible range. Furthermore, the most likely value of each component should be physically reasonable, meaning their centered most likely values shouldn't exceed the component's physical ranges. During this process, ECMs displaying poorly centered distributions or non-physical likely ranges are less preferable and will be deprioritized by AutoEIS. After that, posterior predictive checks are performed by generating simulated EIS data based on the posterior distributions. The metrics used to evaluate their fitting quality are MSE and $R^2$. Here any ECM with an average predictive $R^2$ lower than 0.99 for either the real or imaginary part is regarded as a poor-fitting model and will be deprioritized.

Thirdly, each inference's consistency is assessed using a rank-normalized R-hat diagnostic test ($\hat{R}$), which quantifies the convergence of the Bayesian inference parameters by comparing estimates from independent sampling processes [55] Acceptable $\hat{R}$ values range from 1.00 to 1.05. Any $\hat{R}$ value higher than 1.05 may point out a specification problem and the corresponding ECM will be flagged and assigned with lower priority.

Finally, WAIC values of ECMs are calculated to avoid over-complex models. WAIC evaluates the model's trade-off between fitting quality and model complexity by considering both the log-likelihood and the number of electric components in ECMs. A lower WAIC value is preferable as it indicates a less complex but still sufficiently descriptive model, aligning with the principle of ECM selection. By comparing the WAIC values among ECMs with plausible posteriors distributions and posterior predictions, the ECM with the lowest WAIC is suggested as the correct ECM [56]

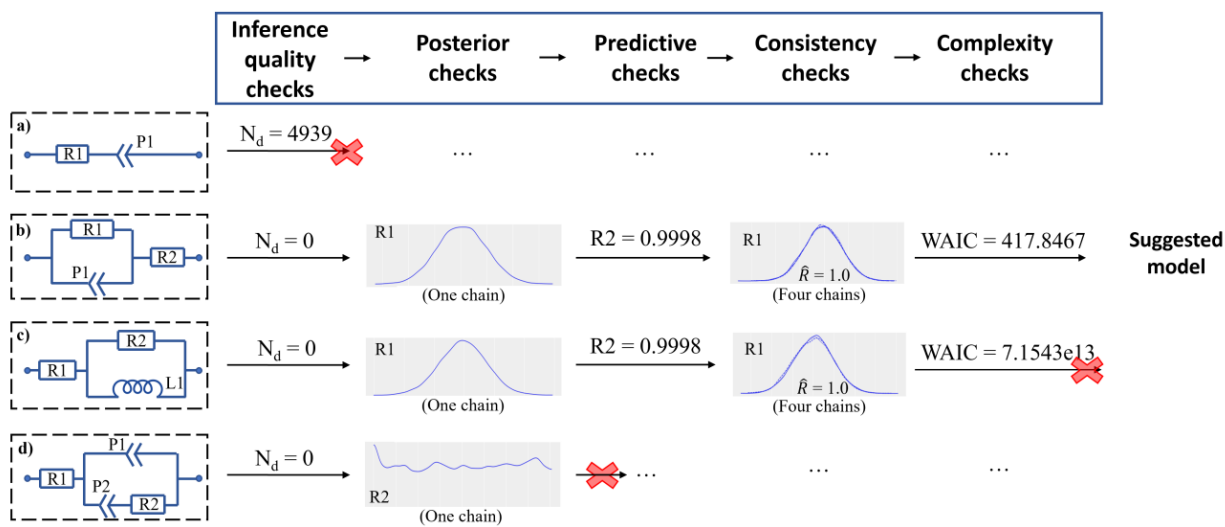

**Figure 2**. The schematic workflow of ECM evaluation

($R$ stands for resistors, $P$ stands for CPEs, and $L$ stands for inductors)

## 3. Results and discussions:

Here we tested AutoEIS with EIS data gathered from three representative electrochemical systems with roughly increasing complexity: OER thin-film electrocatalysts in 3-electrode cells, corrosion of passivating multi-principal element alloys, and $CO_2$ reduction electrolyzer devices.

### 3.1 Oxygen Evolution Reaction Electrocatalysis

OER is an important half-cell reaction in a number of electrocatalytic processes including water splitting and $CO_2$ reduction. The reaction is kinetically sluggish, and typically performed in either acidic or alkaline media. Current state-of-the-art catalysts include iridium or ruthenium oxide (acidic media) or nickel-based oxides (alkaline media) [57]. There is currently a substantial amount of research effort devoted to finding acid or alkaline-stable non-precious metal group catalysts with good catalytic activities and high stabilities [58]. Because of its ability to differentiate multiple interfaces within electrochemical devices, and isolate sluggish electrochemical processes, EIS is regularly used to understand and elucidate the electrochemical processes of OER catalysis [59–61].

The first test case involves the evaluation of electrodeposited thin-film mixed metal oxide catalysts in an acidic environment using a 3-electrode cell configuration. In particular, the catalyst system involves a mildly-stable active catalyst deposited onto a conductive surface, undergoing OER in a highly acidic environment. Given the thin-film nature of the catalyst, it is expected that morphological effects (e.g. porosity) are not present in the system and OER will be the dominant contribution to the EIS [62].

AutoEIS proposes 13 distinct ECMs following the post-filtering process, as depicted in **Figure S1**. All of the ECMs exhibit similar MSE values marginally below 0.01 Ohms$^2$ and $R^2$ values exceeding 0.999. Among the 13 ECMs, 11 contain at least one Randle circuit, while 7 ECMs include inductive components which are not anticipated in this dataset due to the absence of positive imaginary impedances at low frequencies. After performing the Bayesian inference, these non-physical ECMs were effectively deprioritized and the ECM with zero sampling divergences, well-centered posteriors, good consistency, and the lowest WAIC is recommended as the most statistically plausible solution. We report the Bayesian inference results of three representative ECMs from 13 ECMs in **Figure 3**: **(a)** AutoEIS prioritized ECM, **(b)** statistically implausible ECM, and **(c)** the expert's independent solution.

The posterior distributions of each component in AutoEIS's ECM (**Fig 3 (a)**) display Gaussian shapes with reasonably narrow HDI. Also, the inference results of AutoEIS's circuit (**Fig 3 (a)**) exhibit no divergences (marked as black bars), signifying an effective and stable sampling process. In contrast, the statistically implausible circuit (**Fig 3 (b)**) and the expert's independently constructed ECM (**Fig 3 (c)**) both possess imprecise posterior distributions and sampling divergences. For the statistically implausible ECM (**Fig 3 (b)**), the diffuse posterior distribution of R3 and L4 remains the same LogNormal shape and range as their prior distributions. These poorly centered posterior distributions are interpreted as indicating that R3 and L4 have little influence on the quality of the EIS fit, as their HDIs cover multiple orders of magnitude. Consequently, these circuit elements should be considered redundant and removed from the model. Additionally, the inference chain of the statistically implausible ECM (**Fig 3 (b)**) contains numerous divergences, denoting inadequate explorations of its sampling space and the potential existence of implausible circuit structures. Therefore, such an ECM is deprioritized by AutoEIS and should only be used with sufficient physical justification. Another indication of implausible models is the emergence of non-physical most likely value. For the expert circuit (**Fig 3 (c)**), the posterior distribution of P4n concentrates its credibility at the physical upper limit of CPE's alpha term. This might suggest that P4 is better represented as a capacitor, rendering the circuit less preferable. However, the flatness and breadth of the posterior near the upper boundary could also indicate that ascertaining the true value of Pn is challenging due to a lack of sufficient signal. The expert ECM demonstrates a slight improvement in fitting quality (an improvement of 0.0012 in $R^2$) and the WAIC also decreases (from -106.5 to -164.5) as compared to the circuit recommended by AutoEIS. Thus, the addition of the second Randle element does capture more system information without significantly increasing model complexity, at the expense of a less ideal posterior distribution for Pn. AutoEIS explored 100 possible ECMs and identified two reasonable ECMs. An expert would then need to examine and decide upon the appropriate solution for their problem; their analysis follows.

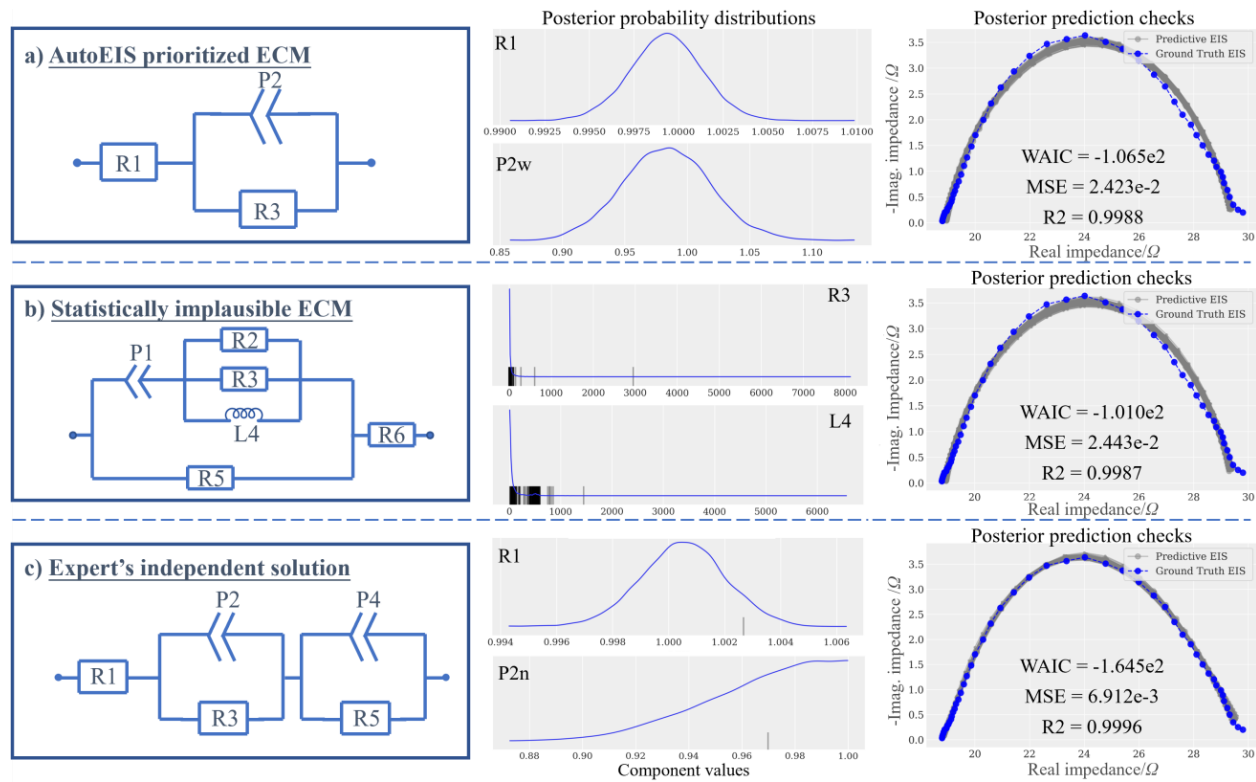

**Figure 3.** The Bayesian inference results of different circuits for the OER measurement **(a)** AutoEIS prioritized ECM **(b)** Statistically implausible ECM **(c)** Expert's independent solution

Given the nature of the electrochemical system being measured, the results of AutoEIS are expected. In particular, the thin-film morphology of the catalyst (minimal porosity) isolates the electrochemical activity to only the OER reaction, with minimal impact on transport processes, which agrees with a single-Randle circuit fit recommended by AutoEIS (i.e., a single electrochemical reaction—OER—dominates the electrochemical response). In particular, from the ECM, we are able to obtain insight into the system ohmic resistance (R1), the charge transfer resistance and the overall kinetic efficiency of the electrochemical reaction (R3), as well as a quantification of the non-ideal capacitance through treatment of the CPE (P2).

The additional complexity of the Expert's independently constructed circuit is based on the intuition of the possibility of re-deposition processes during OER (anodic electrodeposition) which would give rise to a second electrochemical process. However, after discussing with the experts, we reached the consensus this would be a rather small contribution, and is scientifically insignificant for this measurement, as AutoEIS indicates. The results demonstrate AutoEIS successfully identified physically reasonable ECMs for EIS taken on a simple OER catalyst that were consistent with and aided the expert's analysis.

## 3.2 Corrosion of Self-Healing Multi-Principal Element Alloys

The prevention and mitigation of materials corrosion in harsh environments to extend materials' service life are of great priority. Aqueous corrosion, an electrochemical redox reaction involving charge transfer across the metal-electrolyte interface, can be effectively studied using electrochemical techniques such as EIS [63]. Designing self-healing alloys is one of the mitigation routes traditionally followed by using elements such as chromium, and molybdenum [64]. These elements tend to quickly form their oxides/hydroxides under aqueous environments such that a thin passive film forms between the alloy-environment interface, suppressing any further active corrosion. The challenge today is that progress may be nearly exhausted in the study and discovery of binary and ternary alloys with major solutes and one or two impurity elements that control corrosion resistance [65,66]. Such alloys typically contain critical concentrations of key passivating elements. Currently, a great deal of emphasis is directed toward Cantor-type multi-principal element alloys which are argued to be entropy stabilized [67]. These alloys typically have 4 to 6 alloying elements and show indications of unique properties [68]. The corrosion performance and passivation of Cantor alloys are of significant interest and can be controlled by exploring an alloy compositional space that easily numbers in the millions. Impedance has been demonstrated as an effective screening method suitable to rapidly screen numerous compositions. EIS has proven to be an effective tool to probe attributes or parameters functioning to enhance passive film performance in a non-destructive manner [17,18,69].

AutoEIS was used to investigate the EIS response of the steady-state passivation behavior of a homogenized face-centered cubic {FeCoNi}$_{0.84}$-Cr$_{0.16}$ multi-principal element alloy in deaerated solution of 0.1 mol/L H$_2$SO$_{4(aq)}$. The EIS measurements were conducted after passive film was grown for 10 ks at +0.6 V vs. SHE. Four different ECMs were proposed after the post-filtering process. Among the potential ECMs, one simple ECM exhibited poor fitting quality with significant deviation in the low-frequency range while the remaining models were consistent with the given EIS data, as shown in **Figure S3**. **Figure 4** presents the Bayesian inference results of three representative models: **(a)** AutoEIS prioritized ECM, **(b)** statistically implausible ECM, and **(c)** the expert's independent solution.

Upon examining the posterior distributions and predictions of AutoEIS prioritized ECM (**Fig 4 (a)**), all components exhibit narrow and sharp Gaussian posterior distributions without any divergences. In contrast, the inference results of the statistically implausible ECM (**Fig 4 (b)**) and the expert ECM (**Fig 4 (c)**) reveal non-physical most likely values for the ohmic resistance (R1 in the implausible ECM (**Fig 4 (b)**) and R6 in the expert ECM (**Fig 4 (c)**). Further, the R1 and R6 values inferred from the implausible ECM (**Fig 4 (b)**) and the expert ECM (**Fig 4 (c)**) are inconsistent with the ohmic resistance derived from the ground truth EIS data, rendering the circuits less preferable. Furthermore, the posterior distributions of P2n in the statistically implausible ECM (**Fig 4 (b)**) and P1n in the expert ECM (**Fig 4 (c)**) both allocate the majority of their credibility to the physical limit of a CPE (n=1). The sharp increase in credibility and narrow HDI exhibited by the posterior distributions as above indicate the possibility that the CPEs can be replaced by capacitors. Therefore, both the statistically implausible ECM (**Fig 4 (b)**) and the expert

ECM (**Fig 4(c)**) are considered physically less reasonable and deprioritized. As an additional point, comparing the AutoEIS's optimal ECM with the expert's independently constructed ECM, AutoEIS's ECM (**Fig 4 (a)**) shows great similarity to the expert ECM (**Fig 4 (c)**). The Warburg element (W) shown in the expert's circuit is mathematically equivalent to a CPE with an alpha of 0.5, whereas the P2n alpha in AutoEIS's ECM (**Fig 4 (a)**) is 0.42. Although the additional Randle element proposed by experts improves the overall fitting quality without increasing the WAIC, the shape of the posterior distributions dictates that it should be used with caution.

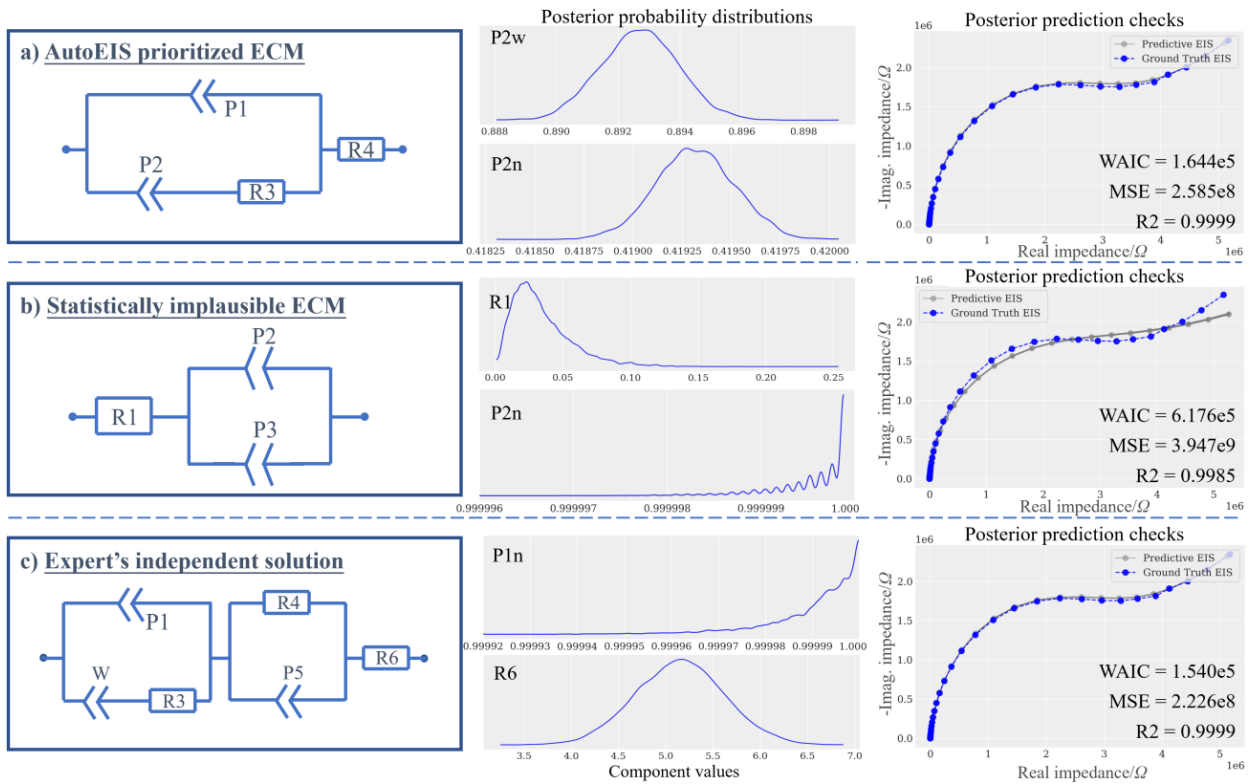

**Figure 4**. The Bayesian inference results of different circuits for the alloy measurement **(a)** AutoEIS prioritized ECM **(b)** Statistically implausible ECM **(c)** Expert's independent solution

AutoEIS's model is physically reasonable given the EIS data. The Randle-like element (shown in the left part in **Fig 4 (a)**) represents the corrosion reaction happening at the defective metal/oxide interface, where P1 represents the capacitive reactance of the interface, P2 represents the impedance offered by the ionic diffusion at low frequencies, and R3 represents the interface resistance. The total ohmic resistance contributed by the electrolyte solution is well represented by R4.

AutoEIS and the expert disagree with regard to the need for an additional Randle element (R4 and P5 in **Fig 4(c)**). The additional element was added due to the expert's expectation that the EIS would include signals from block interfaces created by placing a semiconductor (passive film) in between an electrolyte and metallic alloy. However, in discussions with the expert, the equivalent impedance contribution from the unmatched Randle's element would be orders of magnitude smaller than that of the passive film. Thus, it's scientifically insignificant for measuring passive film polarization resistance, as indicated by AutoEIS. Here, AutoEIS facilitates the exploration of transient film formation and growth for potentially layered passive films in complex concentrated alloys containing multiple passivating elements, exhibiting its ability to assist scientific interpretation of the given EIS data.

### 3.3 $CO_2$ Reduction Electrolyzer Device

Electrochemical conversion of $CO_2$ into useful products such as syn-gas, methane, and ethylene using electro-catalysis is being pursued to mitigate the devastating impact of climate change arising from greenhouse gas emissions [70,71]. However, this electrochemical process suffers from challenges including low selectivity of electro-catalyst, low energy efficiency and low stability due to multiple degradation issues. To improve $CO_2$ electrolysis performance and increase its lifetime stability, it's essential to understand and analyze *in situ* the operation of the $CO_2$ electrolysis cell. EIS is ideal for this purpose due to its ability to dynamically record system changes at different frequencies [72], separating the impact of different processes such as ohmic, charge transfer, and mass transfer losses.

Here we applied AutoEIS to a $CO_2$ electrolyzer device consisting of an anion exchange membrane, silver electro-catalyst as the cathode, and iridium electro-catalyst as the anode. Humidified $CO_2$ was supplied to the cathode, while a liquid electrolyte was fed to the anode. During electrolysis, a mixture of CO and $H_2$ (syn-gas) was generated as products at the cathode, and the OER reaction occurred at the anode. Following the generation and post-filtering of ECMs, AutoEIS proposed 3 candidate ECMs (**Figure S5**). However, AutoEIS did not identify any of them as the optimal ECM during the model evaluation process. Subsequently, we tested the Bayesian inference part of AutoEIS on an expert-constructed ECM and a simplified version of that ECM. The Bayesian inference results of three representative ECMs are reported in **Figure 5**: **(a)** statistically implausible ECM **(b)** the expert's independent solution **(c)** the simplified ECM.

Through the inspection of the posterior distributions of the statistically implausible ECM (**Fig 5 (a)**), it is evident that P4n exhibits a diffuse posterior distribution with the probabilities ranging from 0.6 to 1.0. The broad distribution of P4n suggests that any value of P4n within the range could yield similar fitting results, indicating the potential redundancy of P4. Furthermore, the posterior distribution of R5 in circuit (a) remains identical to its prior distribution in terms of LogNormal shape and range. The poorly centered distribution is interpreted as signifying that R5 has little contribution to the quality of the EIS fit with the consideration of its diffuse distribution across orders of magnitude. Hence, the P4 and R5 are considered redundant, so that the circuit (**Fig 5 (a)**) is deemed statistically implausible and deprioritized by AutoEIS. The expert ECM (**Fig 5 (b)**), constructed using 3 Randle elements, exhibits sharp, narrow posterior distributions of all circuit components, a well-fit predictive plot to the EIS data ($R^2 = 0.9997$), and no divergences. The simplified circuit composed of only two Randle elements was also assessed, since there is always a danger of over-fitting the EIS by using too many Randle elements. Upon applying AutoEIS to the simplified ECM (**Fig 5 (c)**), the posterior distribution of P3w is poorly centered and remains the same shape as its prior, potentially marking the component's redundancy. The HDI of P3n covers the whole physically reasonable range of CPE's alpha term. Such a broad posterior distribution further suggests the implausibility of P3. Moreover, the predictive plot of the simplified ECM (**Fig 5 (c)**) exhibits a larger deviation from the original EIS compared to the expert model (**Fig 5 (b)**) with both lower WAIC and $R^2$ values. This implies that the simplified ECM (**Fig 5 (c)**) is less preferable than the expert circuit (**Fig 5 (b)**). This is indicative of the power the Bayesian Inference contained within AutoEIS's has to discern between ECMs that could be raised during traditional EIS data analysis.

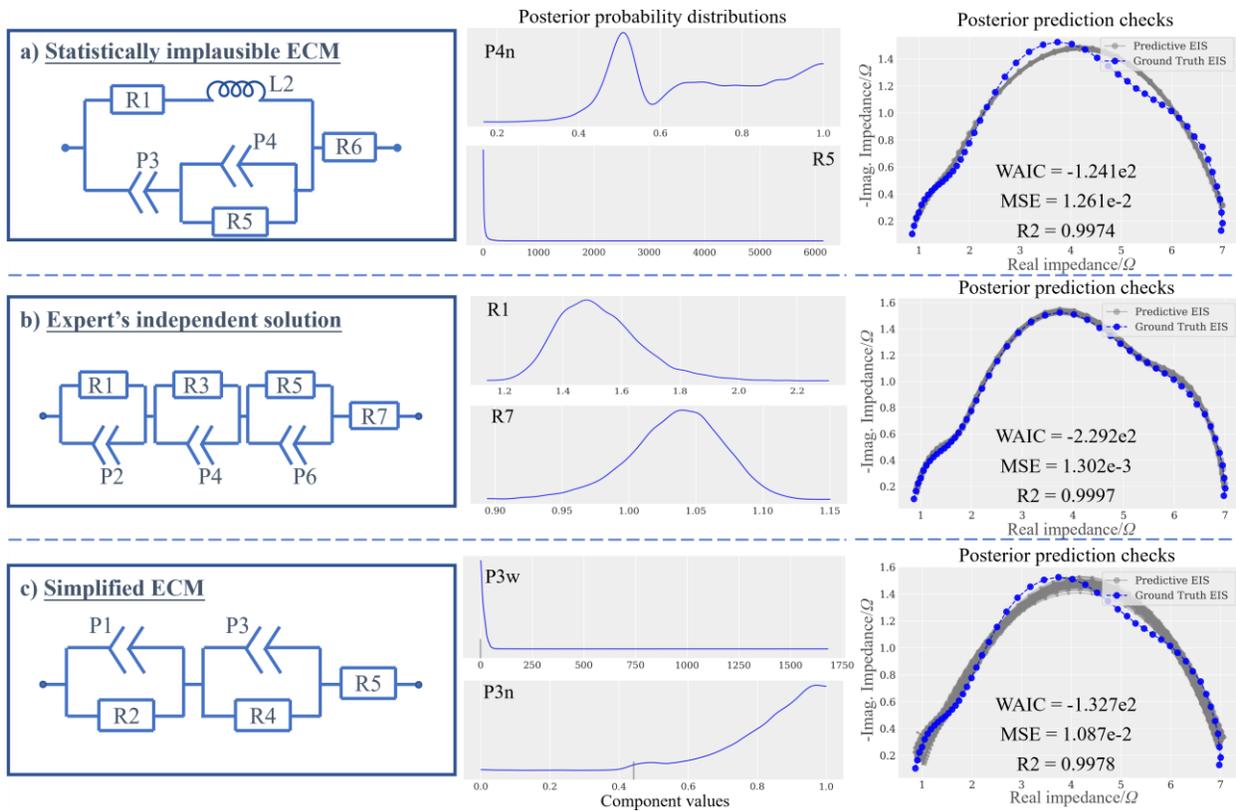

**Figure 5**. The Bayesian inference results of different circuits for the $CO_2$ electrolyzer **(a)** Statistically implausible ECM **(b)** Expert's independent solution **(c)** Simplified ECM

In this system, the evolutionary algorithm contained within AutoEIS was not able to identify a statistically plausible ECM. Looking through the solutions generated, it was observed that there was a tendency to generate nested ECMs with high-level parallel structures as opposed to the more linear series ECMs generated by experts. In principle, the ECM generation could be performed until a statistically plausible circuit is identified, however, this is computationally expensive. Moving forward, it is possible to use supervised machine learning tools to guide the evolutionary algorithm towards ECMs that better reflect expert analysis.

## 4. Conclusion:

This study introduces AutoEIS, a novel tool designed to aid EIS analysis by automatically prioritizing statistically optimal ECM by combining evolutionary algorithms and Bayesian inference. AutoEIS demonstrates great generalizability, enabling expert-level ECM construction without the need for training data or a detailed mechanistic understanding of the electrochemical processes taking place. This allows it to be applied to any electrochemical system in a way that minimizes the impact of human effort and bias in EIS analysis. Furthermore, AutoEIS serves as a valuable tool for distinguishing ECMs with comparable fitting quality, effectively avoiding overfitting through statistical evaluation. We showed through 3 case studies the potential of AutoEIS to enhance and expedite traditional EIS analysis, positioning it as an emerging high-throughput analysis tool for advancing electrochemical materials research. An open-source package is available at GitHub (https://github.com/AUTODIAL/Auto_Eis) and we welcome and encourage community development of the tool.

**Acknowledgment:**

The authors gratefully acknowledge the financial support from Materials for Clean Fuel (MCF) Challenge program at National Research Council of Canada (NRC), the Office of Naval Research (ONR) through the Multidisciplinary University Research Initiative (MURI) program (award #: N00014-20-1-2368) with program manager Dr. Dave Shifler, the National Science Foundation (NSF), and the Material Research Science and Engineering Centers (MRSEC). We also gratefully acknowledge technical discussions and feedback from Dr. Shijing Sun, Prof. Keryn Lian, Dr. Alvin Virya, and Dr. Austin McDannald.

**Conflict of interest:** The authors declare no competing interests.

**Disclaimer:** Certain commercial equipment, instruments, or materials are identified in this paper to facilitate understanding and reproducibility. Such identification does not imply recommendation or endorsement by the National Institute of Standards and Technology, nor does it imply that the materials or equipment identified are necessarily the best available for the purpose.

# AutoEIS: automated Bayesian model selection and analysis for electrochemical impedance spectroscopy

*Runze Zhang[a], Robert Black[b], Debashish Sur[c,d], Parisa Karimi[b], Kangming Li[a], Brian DeCost[e], John R. Scully[c,d], Jason Hattrick-Simpers[a\*]*

- a. Department of Material Science and Engineering, University of Toronto, Toronto, Ontario, Canada

- b. Research Center of Energy, Mining, and Environment, National Research Council Canada, Mississauga, Ontario, Canada

- c. Center for Electrochemical Science and Engineering, University of Virginia, Charlottesville, VA USA 22904

- d. Department of Materials Science and Engineering, University of Virginia, Charlottesville, VA USA 22904

- e. Material Measurement Laboratory, National Institute of Standards and Technology, Gaithersburg, MD, USA

Email: jason.hattrick.simpers@utoronto.ca

**Supporting information:**

1. **Oxygen Evolution Reaction Electrocatalysis**

The 13 different ECMs proposed by AutoEIS after the filtering process are shown below in **Figure S1**. It can be observed that all the ECMs generated by the evolutionary algorithm exhibit great consistency with the original data and the fitting quality between different ECMs are very close, hence it is difficult to distinguish them by using only the $R^2$ value or MSE metric.

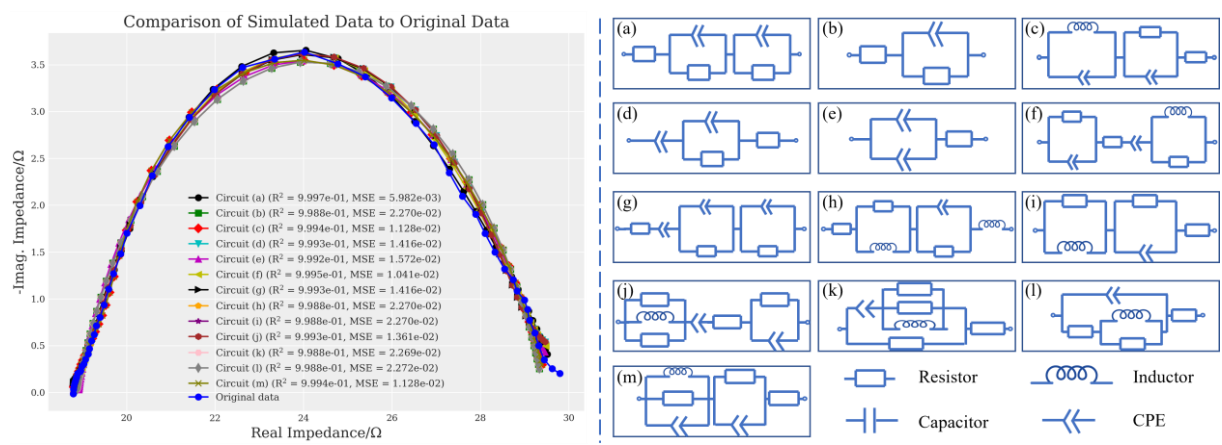

**Figure S1.** The fitting qualities of ECMs on the OER measurement

The detailed Bayesian inference results for the three representative models for the OER measurement reported in the main text are shown below in Figure S2: (a) AutoEIS prioritized ECM (b) Statistically implausible ECM (c) the Expert's independent solution.

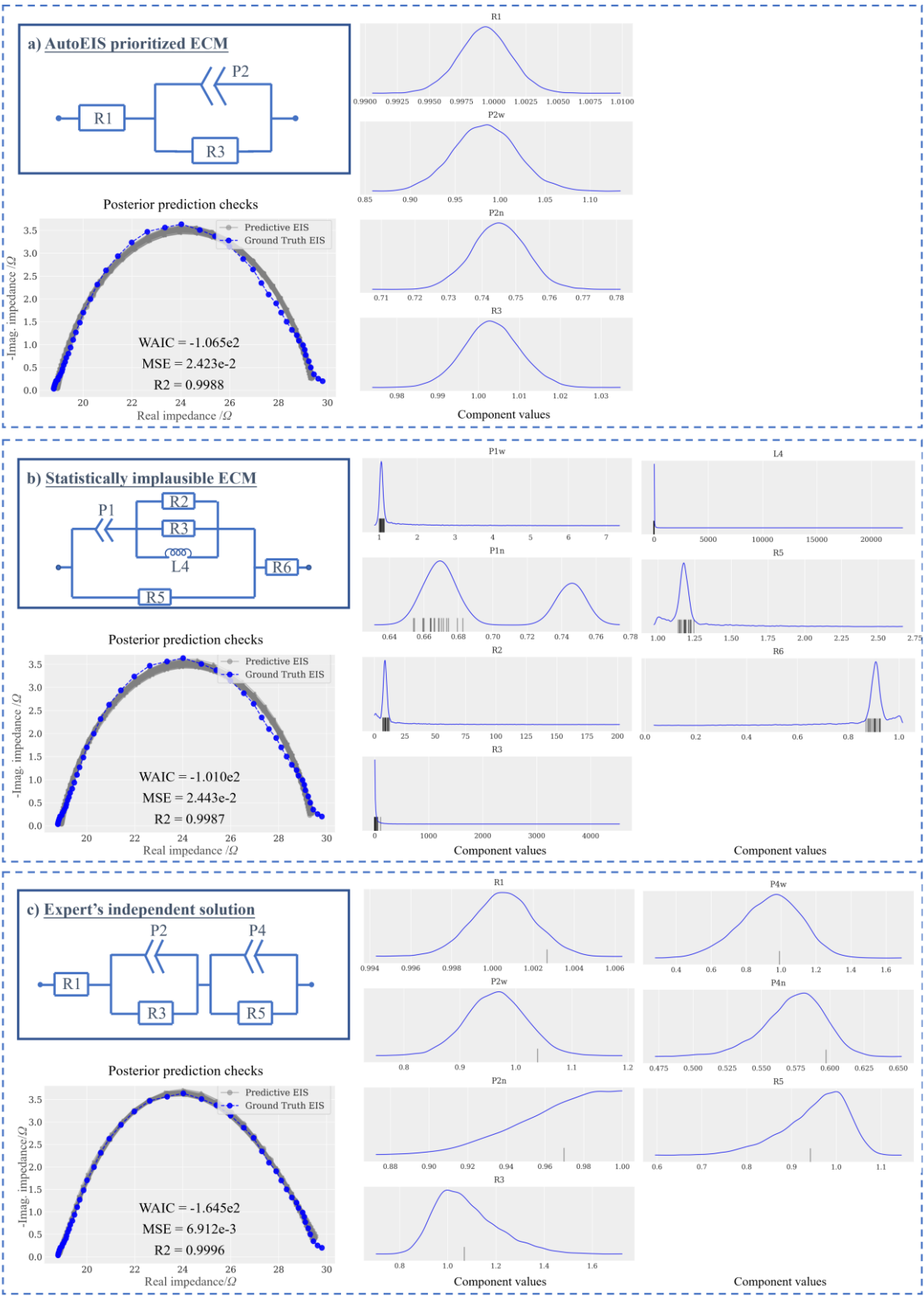

**Figure S2**. The Bayesian inference results of different circuits for the OER measurement **(a)** AutoEIS prioritized ECM **(b)** Statistically implausible ECM **(c)** Expert's independent solution

## 2. Corrosion of Self-Healing Multi-Principal Element Alloys

The 4 different ECMs proposed by AutoEIS after the filtering process are shown below in **Figure S3**. It can be observed that only circuit (b) showed obvious deviation from the original data in the low-frequency range, while the rest three ECMs are consistent with the given EIS data.

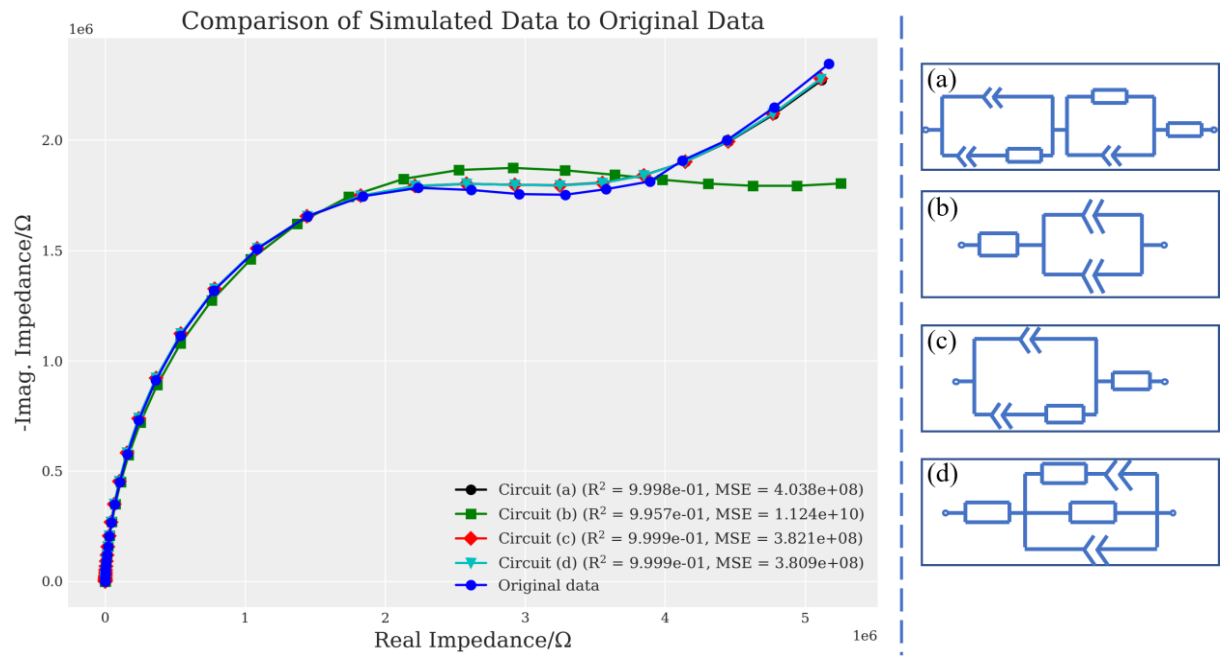

**Figure S3.** The fitting qualities of ECMs on the alloy measurement

The detailed Bayesian inference results for the three representative models for the alloy measurement reported in the main text are shown below in **Figure S4**: **(a)** AutoEIS prioritized ECM **(b)** Statistically implausible ECM **(c)** the Expert's independent solution.

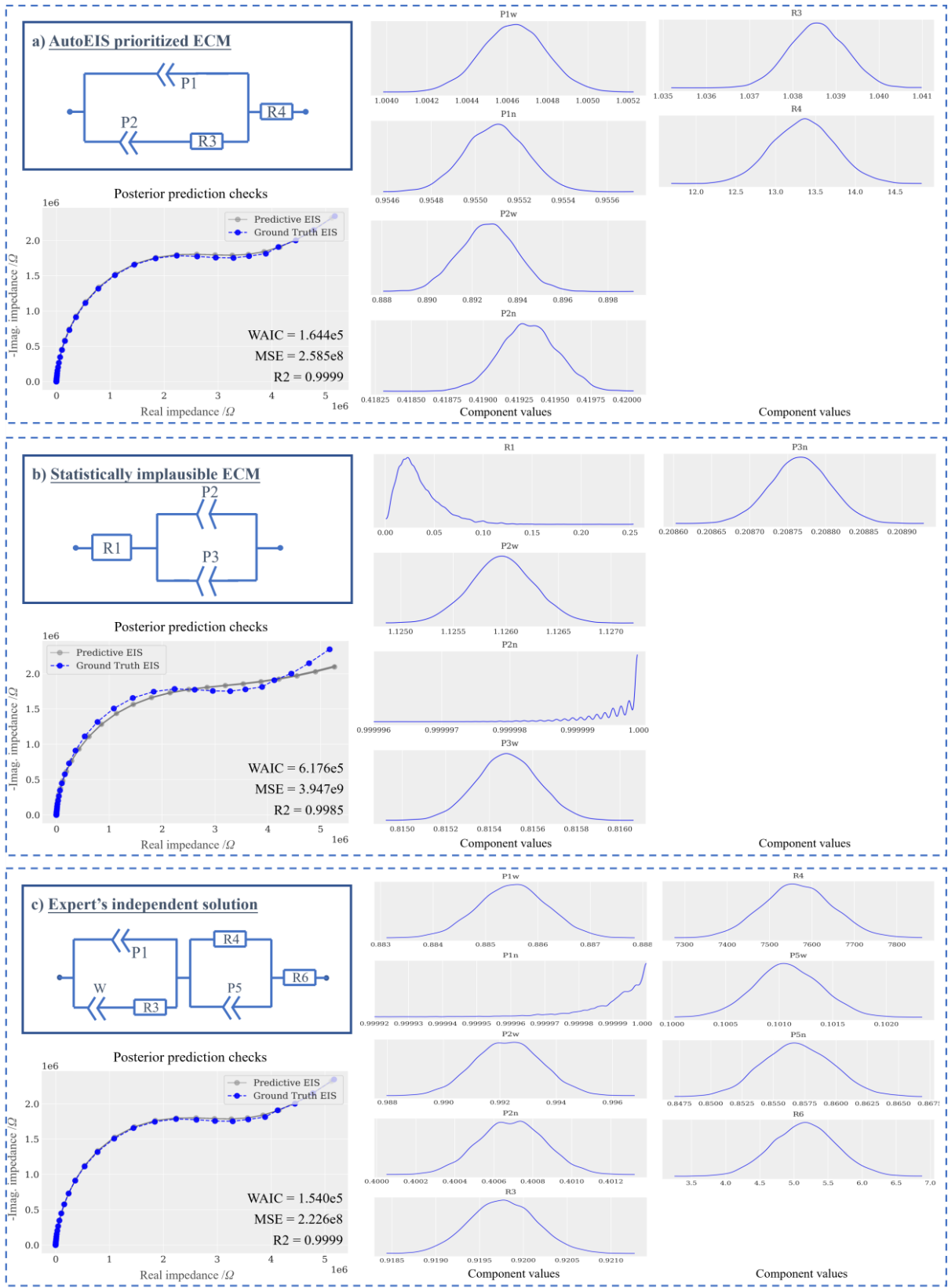

**Figure S4**. The Bayesian inference results of different circuits for the alloy measurement

**(a)** AutoEIS prioritized ECM **(b)** Statistically implausible ECM **(c)** Expert's independent solution

### 3. $CO_2$ Reduction Electrolyzer Device

The 3 different ECMs proposed by AutoEIS after the filtering process are shown below in **Figure S5**. It can be observed that the ECMs raised by the evolutionary algorithm didn't achieve a very good fit on the given ECM data.

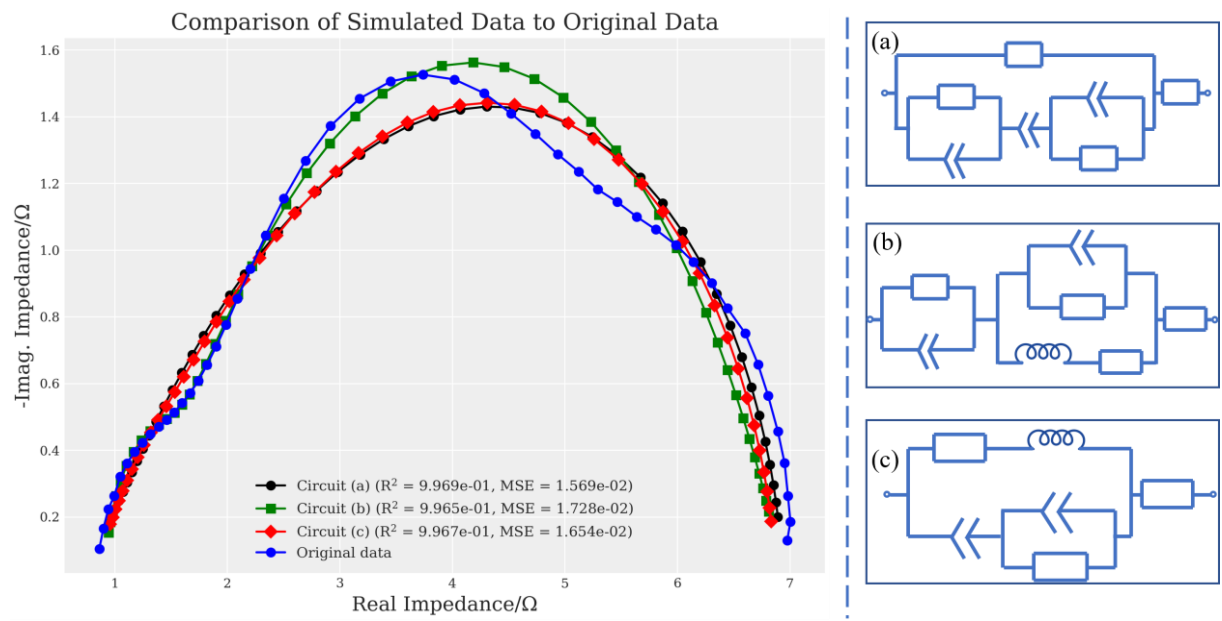

**Figure S5.** The fitting qualities of ECMs on the $CO_2$ electrolyzer measurement

The detailed Bayesian inference results for the three representative models for the $CO_2$ electrolyzer measurement reported in the main text are shown below in **Figure S6**: **(a)** Statistically implausible ECM **(b)** the Expert's independent solution **(c)** the Simplified ECM

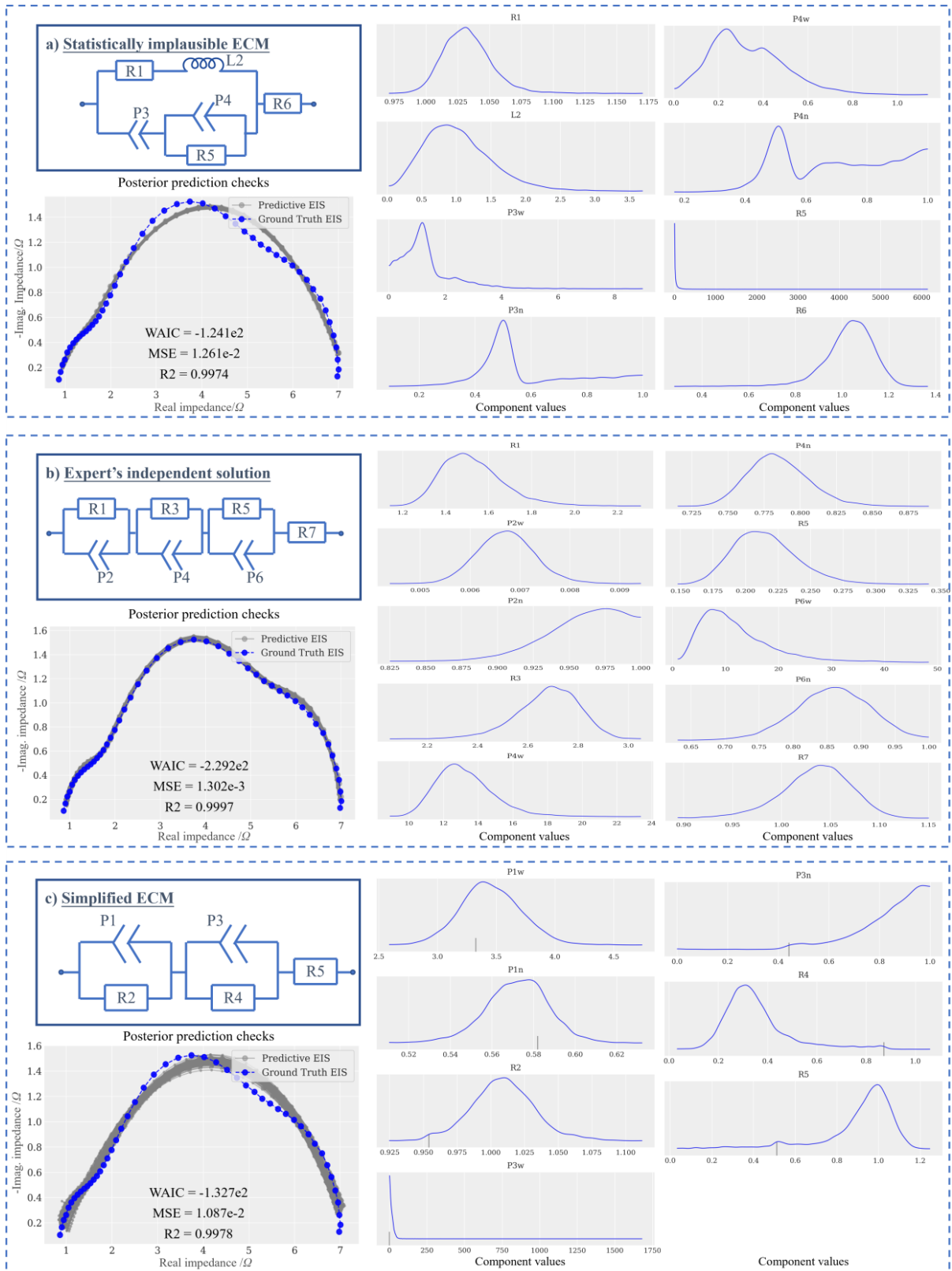

**Figure S6**. The Bayesian inference results of different circuits for the $CO_2$ electrolyzer measurement

**(a)** Statistically implausible ECM **(b)** Expert's independent solution **(c)** Simplified ECM